# Structural Properties of an S-system Model of *Mycobacterium Tuberculosis* Gene Regulation

**Honeylou F. Farinas**[1,2], **Eduardo R. Mendoza**[2,3,4,5], and **Angelyn R. Lao**[2*]

[1]Department of Mathematics, Mariano Marcos State University, Ilocos Norte, 2906 Philippines

[2]Mathematics and Statistics Department, De La Salle University, Manila, Metro Manila 1004 Philippines

[3]Institute of Mathematical Sciences and Physics, University of the Philippines, Los Banos, Laguna 4031 Philippines

[4]Max Planck Institute of Biochemistry, Martinsried near Munich Germany

[5]Faculty of Physics, Ludwig Maximilian University, Munich 80539, Germany

ABSTRACT

Magombedze and Mulder (2013) studied the gene regulatory system of *Mycobacterium Tuberculosis* (*Mtb*) by partitioning this into three subsystems based on putative gene function and role in dormancy/latency development. Each subsystem, in the form of S-system, is represented by an embedded chemical reaction network (CRN), defined by a species subset and a reaction subset induced by the set of digraph vertices of the subsystem. For the embedded networks of S-system, we showed interesting structural properties and proved that all S-system CRNs (with at least two species) are discordant. Analyzing the subsystems as subnetworks, where arcs between vertices belonging to different subsystems are retained, we formed a digraph homomorphism from the corresponding subnetworks to the embedded networks. Lastly, we explored the modularity concept of CRN in the context of digraph.

## 1. INTRODUCTION

Tuberculosis (TB) is one of the oldest infectious diseases affecting mankind caused by *Mycobacterium tuberculosis* (*Mtb*). It is an unusual bacterial pathogen which has the remarkable ability to cause both acute-life threatening disease and also clinically latent infections which can persist for the lifetime of the human host (Young et al. 2008). About a quarter of the world's population is infected with *Mtb,* and in 2018 an estimated 1.5 million TB deaths were reported (WHO 2019).

To advance understanding and treatment of TB, mathematical and computational models are used to integrate data across multiple time and length scales. A useful classification of TB models was introduced by Kirschner et al. in their review paper (Kirschner 2017). The first group consists of epidemiological models, which describe the population-level dynamics of TB under different scenarios. The second group explores various aspects of *Mtb* as an independent microorganism such as growth, drug resistance development, metabolism and adaptive response under various conditions and the last focuses on within-host TB models.

In this paper, we pay attention to two S-system models of the gene regulatory system of *Mtb* developed by Magombedze & Mulder (2013), which belong to the second classification of TB models. S-system models are typically derived from biochemical maps (Voit 2000), which are based on digraphs whose vertices and arcs represent molecules and interactions of a biochemical system.

The two S-system models correspond to two phases of the pathogen's life cycle: the Non-Replicating Phase (NRP) and the Stationary Phase (STR). The NRP model simulates the development of dormant *Mtb* by gradual oxygen depletion over 80 days. While the STR model simulates latent *Mtb* over 60 days. The gene regulatory system of *Mtb* was formalized as digraphs (see Figure 1), with the vertices representing the genes and the arcs representing the interactions between them.

A vertex set partition, based on putative gene function and role in dormancy/latency development, defines three subdigraphs in both models: $\mathcal{D}_1, \mathcal{D}_2$ and $\mathcal{D}_3$. Subdigraph $\mathcal{D}_1$, is constituted by genes that are involved in the bacilli's cell wall process. Genes that have functions related to the DosR-regulon and adaptation are clustered in $\mathcal{D}_3$ and several genes that have other functions constitute $\mathcal{D}_2$. The arcs between vertices that do not belong to the same subdigraphs are deleted.

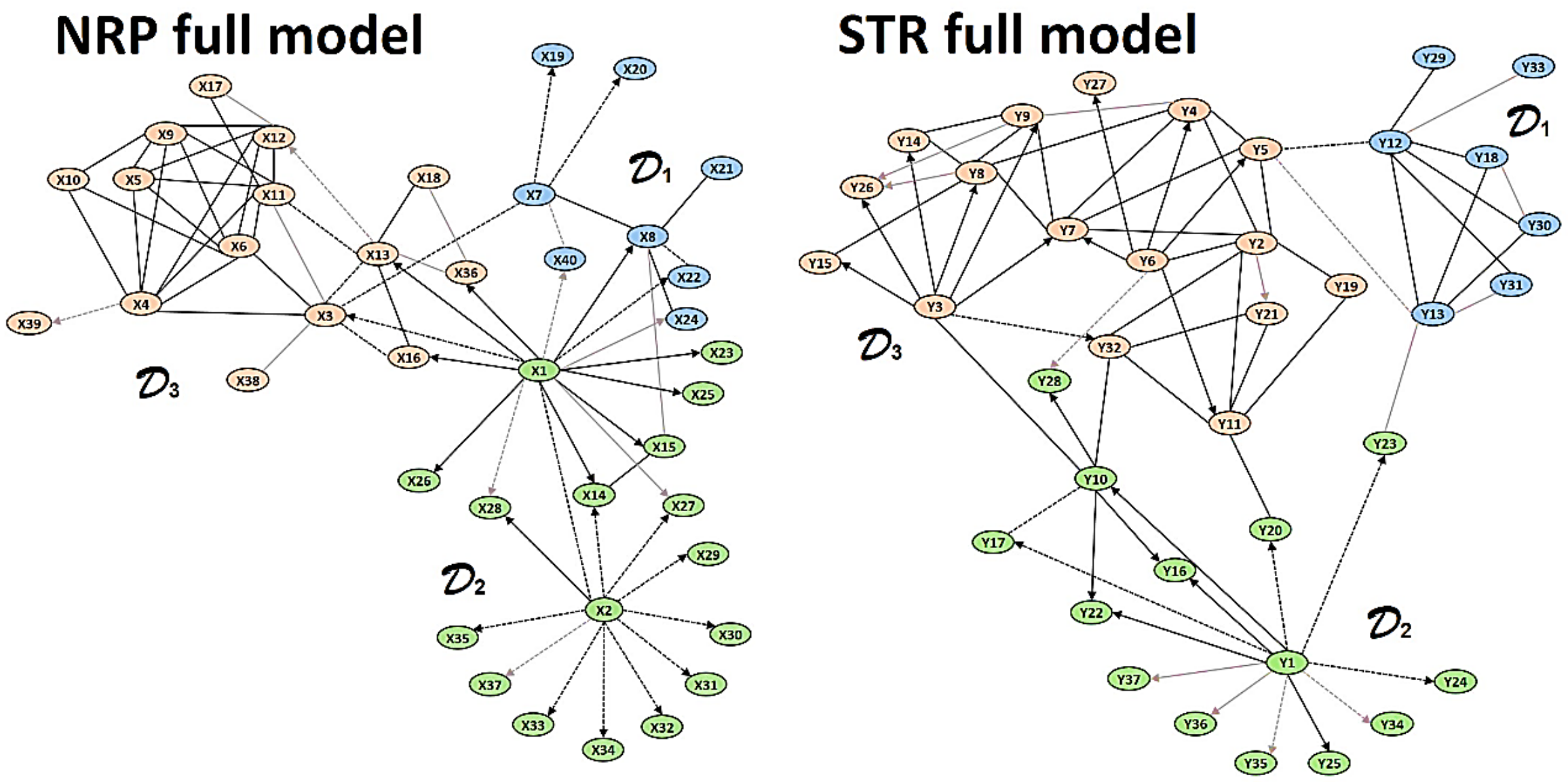

**Figure 1.** Digraphs of the NRP and STR models. The vertices represent the genes and the arcs represent the interactions between them. Arcs with no direction indicate bi-directional interaction. Both digraphs share a number of similar genes (See the Supplementary Information in Magombedze & Mulder (2013) for the corresponding labels of the genes.)

An S-system – with the digraph as its "biochemical map" – is then built for both NRP (left side of Figure 1) and STR (right side of Figure 1) models with the genes (vertices of the digraph) as variables. The models assumed that gene interactions will result in the change of expression of one gene, in such a way that all interactions should contribute to the term which is modelled as a function responsible for gene regulation in the system. Thus, the S-systems for NRP and STR considered the genes as the dependent variables $X_i$ and no independent variables. It is represented as

$$\frac{dX_i}{dt} = \alpha_i \prod_{j=1}^{m} X_j^{g_{ij}} - \beta_i X_i \ , \quad i \in 1, \dots , m$$

where the constants $\alpha_i$ and $\beta_i$ are nonnegative and the exponent $g_{ij}$ is real. The S-system ODEs for the NRP and STR models can be found in the Supplementary Information of Magombedze & Mulder (2013). Modularity of the system is often expressed by partitioning the vertices into subsets and forming corresponding subsystems. The subsystems of the S-system correspond to the subdigraphs $\mathcal{D}_1$, $\mathcal{D}_2$ and $\mathcal{D}_3$ of the NRP and STR models.

In this paper, we use Chemical Reaction Network (CRN) representation of an S-system introduced by Arceo et al. (2015) to represent the corresponding NRP and STR S-systems and investigate interesting structural properties of the network. Chemical Reaction Network Theory (CRNT) analysis of the *Mtb* S-systems shed light in understanding the structural properties of the networks formed by the subsystems. However, we did not relate the CRN representations to the dynamical behavior of the *Mtb*

S-systems due to the extensive dynamic analysis of Magombedze and Mulder in their paper (Magombedze & Mulder 2013).

Our analysis of the *Mtb* S-systems is two-fold: each subsystem is represented as embedded networks (an arc is deleted between a vertex connecting the genes into different subsystems) and as subnetworks (an arc is retained between a vertex belonging to different subsystems).

The main results of this paper are:

1. We show that the structural properties of the S-system model's CRNs are identical to those of the BST models studied in Arceo et al. (2015), as indicated in Table 2.
2. To establish the above claim, we
   - derive general relationships between several structural properties (Proposition 2 and Theorem 1)
   - prove that, in general, the CRN of an S-system with two or more dependent variables is discordant (Theorem 3)
3. Since the modularity of the S-system's CRN – considered as a digraph – is inconsistent with the modularity of the original digraph, we propose a novel concept (Definition 22) that incorporates stoichiometric properties into measures of a CRN's modularity (Table 5).

In addition, we show that there are surjective digraph homomorphisms from the subnetworks to the corresponding embedded networks of the S-system CRN, which could be useful for studying further structural properties of the system.

## 2. FUNDAMENTALS OF CHEMICAL REACTION NETWORKS AND KINETIC SYSTEMS

In this section, we introduce some fundamental concepts of CRNT presented in Arceo et al. (2015), Feinberg (1987) and Fortun et al. (2018). The mathematical notations used throughout the paper were also defined here.

**Definition 1.** A **chemical reaction network** $\mathcal{N} = (\mathcal{S}, \mathcal{C}, \mathcal{R})$ is a triple of three non-empty finite sets:

1. A set **species** $\mathcal{S}$
2. A set $\mathcal{C}$ of **complexes**, which are non-negative integer linear combinations of the species, and
3. A set $\mathcal{R} \subseteq \mathcal{C} \times \mathcal{C}$ of **reactions** such that
   - $(i, i) \notin \mathcal{R}$ for all $i \in \mathcal{C}$, and

- For each $i \in \mathcal{C}$, there exists a $j \in \mathcal{C}$ such that $(i, j) \in \mathcal{R}$ or $(j, i) \in \mathcal{R}$.

We denote with $\boldsymbol{m}$ the number of species, $\boldsymbol{n}$ the number of complexes and $\boldsymbol{r}$ the number of reactions in a CRN.

**Definition 2.** A complex is called **monospecies** if it consists of only one species, i.e. of the form $kX_i$, $k$ a non-negative integer and $X_i$ a species. It is called **monomolecular** if $k = 1$, and is identified with the **zero complex** for $k = 0$.

A zero complex represents the "outside" of the system studied, from which chemicals can flow into the system at a constant rate and to which they can flow out at a linear rate (proportional to the concentration of the species). In biological systems, the "outside" also stands for the degradation of a species.

A CRN $\mathcal{N} = (\mathcal{S}, \mathcal{C}, \mathcal{R})$ gives rise to a digraph with complexes as vertices and reactions as arcs. However, the digraph determines the triple uniquely only if an additional property is considered in the definition: $\mathcal{S} = \cup \operatorname{supp} i$ for $i \in \mathcal{C}$, i.e., each species appears in at least one complex. With this additional property, a CRN can be equivalently defined as follows:

**Definition 3.** A **chemical reaction network** is a digraph ($\mathcal{C}$,$\mathcal{R}$) where each vertex has positive degree and stoichiometry, i.e., there is a finite set $\mathcal{S}$ (whose elements are called **species**) such that $\mathcal{C}$ is a subset of $\mathbb{R}^{\mathcal{S}}_{\geq}$. Each vertex is called a **complex** and its coordinates in $\mathbb{R}^{\mathcal{S}}_{\geq}$ are called **stoichiometric coefficients**. The arcs are called **reactions**.

Two useful maps are associated with each reaction:

**Definition 4.** The **reactant map** $\rho : \mathcal{R} \to \mathcal{C}$ maps a reaction to its reactant complex while the **product map** $\pi : \mathcal{R} \to \mathcal{C}$ maps it to its product complex. We denote $|\rho(\mathcal{R})|$ with $n_r$, i.e., the number of reactant complexes.

Connectivity concepts in Digraph Theory apply to CRNs, but have slightly differing names. A connected component is traditionally called a **linkage class**, denoted by $\mathcal{L}$, in CRNT. A subset of a linkage class where any two elements are connected by a directed path in each direction is known as a **strong linkage class**. If there is no reaction from a complex in the strong linkage class to a complex outside the same strong linkage class, then we have a **terminal strong linkage class**. We denote the number of linkage classes with $\boldsymbol{l}$, that of the strong linkage classes with $\boldsymbol{sl}$ and that of terminal strong linkage classes with $\boldsymbol{t}$. Clearly, $sl \geq t \geq l$.

Many features of CRNs can be examined by working in terms of finite dimensional spaces $\mathbb{R}^{\mathcal{S}}$, $\mathbb{R}^{\mathcal{C}}$, $\mathbb{R}^{\mathcal{R}}$, which are referred to as species space, complex space and reaction space, respectively. We can view a complex $j \in \mathcal{C}$ as a vector in $\mathbb{R}^{\mathcal{C}}$ (called *complex vector*) by writing $j = \sum_{s \in \mathcal{S}} j_s s$, where $j_s$ is the stoichiometric coefficient of species $s$.

**Definition 5.** The **reaction vectors** of a CRN $\mathcal{N} = (\mathcal{S}, \mathcal{C}, \mathcal{R})$ are the members of the set $\{j - i \in \mathbb{R}^{\mathcal{S}} | (i, j) \in \mathcal{R}\}$. The **stoichiometric subspace** $S$ of the CRN is the linear subspace of $\mathbb{R}^{\mathcal{S}}$ defined by $S : span\,\{j - i \in \mathbb{R}^{\mathcal{S}} \,|(i, j) \in \mathcal{R}\}$. The **rank** of the CRN, $\boldsymbol{s}$**,** is defined as $s = \dim S$.

**Definition 6.** The **incidence map** $I_a : \mathbb{R}^{\mathcal{R}} \to \mathbb{R}^{\mathcal{C}}$ is defined as follows: For $f : \mathcal{R} \to \mathbb{R}$, then $I_a(f)(v) = -f(a)$ and $f(a)$ if $v = \rho(a)$ and $v = \pi(a)$, respectively, and are $0$ otherwise.

Equivalently, it maps the basis vector $\omega_a$ to $\omega_{v'} - \omega_v$ if $a: v \to v'$. It is clearly a linear map, and its matrix representation (with respect to the standard bases $\omega_a, \omega_v$) is called the **incidence matrix**, which can be described as

$$(I_a)_{i,j} = \begin{cases} -1 & \text{if } \rho(a_j) = v_i, \\ 1 & \text{if } \pi(a_j) = v_i, \\ 0 & \text{otherwise.} \end{cases}$$

Note that in most digraph theory books, the incidence matrix is set as $-I_a$. An important result of digraph theory regarding the incidence matrix is the following:

**Proposition 1.** Let $I$ be the incidence matrix of the directed graph $D = (V, E)$. Then rank $I = n - l$, where $l$ is the number of connected components of $D$.

A non-negative integer, called the deficiency, can be associated to each CRN. This number has been the center of many studies in CRNT due to its relevance in the dynamic behavior of the system.

**Definition 7.** The **deficiency** of a CRN is the integer $\boldsymbol{\delta} = n - l - s$.

**Definition 8.** The **reactant subspace** $R$ is the linear space in $\mathbb{R}^{\mathcal{S}}$ generated by the reactant complexes. Its dimension, $\dim R$ denoted by $\boldsymbol{q}$, is called the **reactant rank** of the network. Meanwhile, the **reactant deficiency** $\boldsymbol{\delta_p}$ is the difference between the number of reactant complexes and the reactant rank, i.e., $\delta_p = n_r - q$.

We now introduce the fundamentals of chemical kinetic systems. We begin with the general definition of kinetics from Feliu & Wiuf (2012):

**Definition 9.** A **kinetics** for a CRN $(\mathcal{S}, \mathcal{C}, \mathcal{R})$ is an assignment of a rate function $K_j : \Omega_K \to \mathbb{R}_>$ to each reaction $r_j \in \mathcal{R}$, where $\Omega_K$ is a set such that $\mathbb{R}^{\mathcal{S}}_> \subseteq \Omega_K \subseteq \mathbb{R}^{\mathcal{S}}_\geq, c \wedge d \in \Omega_K$ whenever $c, d \in \Omega_K$, and $K_j(c) \geq 0,\ \forall c \in \Omega_K$. A kinetics for a network $\mathcal{N}$ is denoted by $K = (K_1, K_2, \ldots, K_r)$: $\Omega_K \to \mathbb{R}^{\mathcal{R}}_\geq$. The pair $(\mathcal{N}, K)$ is called the **chemical kinetic system** (CKS).

In the definition, $c \wedge d$ is the bivector of $c$ and $d$ in the exterior algebra of $\mathbb{R}^{\mathcal{S}}$. We add the definition relevant to our context:

**Definition 10.** A **chemical kinetics** is a kinetics $K$ satisfying the positivity condition: for each reaction $j : y \to y'$, $K_j(c) > 0$ iff supp $y \subset$ supp $c$.

**Definition 11.** The **species formation rate function** (SFRF) of a Chemical Kinetic System (CKS) is the vector field $f(x) = NK(x) = \sum_{y \to y'} K_{y \to y'}(x)(y' - y)$

Once a kinetics is associated with a CRN, we can determine the rate at which the concentration of each species evolves at composition $c$.

Power-law kinetics is defined by an $r\ x\ m$ matrix $F = [F_{ij}]$, called the **kinetic order matrix**, and vector $k \in \mathbb{R}^{\mathcal{R}}$, called the **rate vector**. In power-law formalism, the kinetic orders of the species concentrations are real numbers.

**Definition 12.** A kinetics $K: \mathbb{R}^{\mathcal{R}}_{>} \to \mathbb{R}^{\mathcal{R}}$ is a **power-law kinetics** (PLK) if $K_i(x) = k_i x^{F_i}\ \ \forall i = 1, \ldots, r$ with $k_i \in \mathbb{R}_{>}$ and $F_{i,j} \in \mathbb{R}$.

A subnetwork and an embedded network of a CRN were introduced by Joshi  and Shiu (Joshi & Shiu 2012; Joshi & Shiu 2013). The definitions are based on the concept of restriction maps between subsets in a network's sets.

**Definition 13.** Let $(\mathcal{S}, \mathcal{C}, \mathcal{R})$ be a CRN. Consider a subset of the species $S \subset \mathcal{S}$, a subset of the complexes $C \subset \mathcal{C}$ and a subset of the reactions $R \subset \mathcal{R}$.

1. The **restriction of** $R$ **to** $S$, denoted by $R|_S$, is the set of reactions obtained by taking the reactions in $R$ and removing all species not in $S$ from the reactant and product complexes. If a reactant or a product complex does not contain any species from the set $S$, then the complex is replaced by the 0 complex in $R|_S$. If a trivial reaction (one in which the reactant and product complexes are the same) is obtained in this process, then that reaction is removed. Also, extra copies of repeated reactions are removed.

2. The **restriction of** $C$ **to** $R$, denoted by $C|_R$, is the set of (reactant and product) complexes of the reactions in $R$.

3. The **restriction of** $S$ **to** $C$, denoted by $S|_C$, is the set of species that are in the complexes in $C$.

**Definition 14.** Let $(\mathcal{S}, \mathcal{C}, \mathcal{R})$ be a CRN.

1. A subset of the reactions $\mathcal{R}' \subset \mathcal{R}$ defines a **subnetwork** $(\mathcal{S}|_{\mathcal{C}|\mathcal{R}'}, \mathcal{C}|_{\mathcal{R}'}, \mathcal{R}')$, where $\mathcal{C}|_{\mathcal{R}'}$ denotes the set of complexes that appear in the reactions $\mathcal{R}'$ and $\mathcal{S}|_{\mathcal{C}|\mathcal{R}'}$ denotes the set of species that appear in those complexes.
2. An **embedded network** of $(\mathcal{S}, \mathcal{C}, \mathcal{R})$, which is defined by a subset of the species, $S = \{X_{i1}, X_{i2}, \dots, X_{ik}\} \subset \mathcal{S}$, and a subset of the reactions $R = \{R_{j1}, R_{j2}, \dots, R_{jl}\} \subset \mathcal{R}$, that involve all species of $S$, is the network $(S, \mathcal{C}|_{R|S}, R|_S)$ consisting of the reactions $R|_S$.

In the context of this paper, the subnetworks and embedded networks of the CRN representation of an $S$-system are not necessarily the same. We used $\mathcal{N}_i$ and $\mathcal{N}_i^*$ to denote the subnetwork and embedded network of a CRN $\mathcal{N}$, respectively. The set of subnetworks $\mathcal{N}_i$ is obtained by removing a subset of reactions, while an embedded network $\mathcal{N}_i^*$ is obtained by removing a subset of reactions or subsets of species or both. For instance, removing the species $X_2$ from the reaction $X_1 + X_2 \rightarrow X_1 + X_3$ results in the reaction $X_1 \rightarrow X_1 + X_3$.

In some cases, removing species results in a trivial reaction – where the source and product complex are identical. For instance, removal of both $X_2$ and $X_3$ from $X_1 + X_2 \rightarrow X_1 + X_3$ results in the trivial reaction $X_1 \rightarrow X_1$. So, after removing species, any trivial reactions and any copies of duplicate reactions are discarded.

Feinberg's general concept of a network decomposition is as follows:

**Definition 15.** A set of subnetworks $\mathcal{N}_i = (\mathcal{S}_i, \mathcal{C}_i, \mathcal{R}_i)$ is a **network decomposition** of $\mathcal{N} = (\mathcal{S}, \mathcal{C}, \mathcal{R})$ if $\{\mathcal{R}_i\}$ forms a partition of $\mathcal{R}$.

Note that partitioning the reaction set does not necessarily partition the set of complexes. A basic property of decomposition is that $s \leq \sum s_i$ where $s$ and $s_i$ are the dimensions of the stoichiometric subspaces of the network $\mathcal{N}$ and its corresponding subnetworks $\mathcal{N}_i$, respectively. Meanwhile, if they are equal, then the network is **independent** (Feinberg 1987).

The best known and most studied decomposition is the linkage class decomposition. Independence of linkage classes (ILC) is an important feature of a CRN. Boros (2013) showed that a necessary and sufficient condition for a linkage class to be independent is $\delta = \sum \delta_i$. Generally, if a network has an independent decomposition then $\delta \leq \sum \delta_i$ (Fortun et al. 2018).

## 3. PROPERTIES OF THE S-SYSTEM AND THEIR EMBEDDED NETWORKS

The CRNs of the NRP and STR models, as well as the three embedded networks in each model, are documented in Appendix I. Table 1 summarizes the network numbers for the CRN representations of the S-system of the NRP and STR full models $\mathcal{N}$ and their corresponding embedded networks $\mathcal{N}_1^*, \mathcal{N}_2^*$ and $\mathcal{N}_3^*$.

**Table 1.** NRP and STR network numbers for $\mathcal{N}$ and its embedded networks $\mathcal{N}_1^*, \mathcal{N}_2^*$ and $\mathcal{N}_3^*$.

| Network numbers | NRP | | | | STR | | | |
|---|---|---|---|---|---|---|---|---|
| | $\boldsymbol{\mathcal{N}}$ | $\boldsymbol{\mathcal{N}_1^*}$ | $\mathcal{N}_2^*$ | $\mathcal{N}_3^*$ | $\boldsymbol{\mathcal{N}}$ | $\boldsymbol{\mathcal{N}_1^*}$ | $\mathcal{N}_2^*$ | $\mathcal{N}_3^*$ |
| Species ($\boldsymbol{m}$) | 40 | 8 | 17 | 15 | 37 | 7 | 14 | 16 |
| Irreversible species ($\boldsymbol{m_{rev}}$) | 0 | 0 | 0 | 0 | 1 | 0 | 1 | 1 |
| Complex ($\boldsymbol{n}$) | 98 | 19 | 36 | 42 | 95 | 19 | 29 | 44 |
| Source complex ($\boldsymbol{n_r}$) | 60 | 10 | 20 | 28 | 60 | 12 | 17 | 29 |
| Irreversible reaction ($\boldsymbol{r_{irrev}}$) | 80 | 16 | 34 | 30 | 72 | 14 | 26 | 30 |
| Linkage class ($\boldsymbol{l}$) | 19 | 3 | 3 | 12 | 22 | 5 | 2 | 13 |
| Strong linkage class ($\boldsymbol{sl}$) | 98 | 19 | 36 | 42 | 94 | 19 | 28 | 43 |
| Terminal linkage class ($\boldsymbol{t}$) | 38 | 9 | 16 | 14 | 35 | 7 | 12 | 15 |
| Rank ($\boldsymbol{s}$) | 40 | 8 | 17 | 15 | 37 | 7 | 14 | 16 |
| Reactant rank ($\boldsymbol{q}$) | 40 | 8 | 17 | 15 | 37 | 7 | 14 | 16 |
| Deficiency ($\boldsymbol{\delta}$) | 39 | 8 | 16 | 15 | 36 | 7 | 13 | 15 |
| Reactant deficiency ($\boldsymbol{\delta_p}$) | 20 | 2 | 3 | 13 | 23 | 5 | 3 | 13 |

### 3.1 Network Properties of the NRP and STR Models

The NRP and STR full models and their embedded networks are fully open networks since each species has an outflow reaction as seen in their corresponding CRN in Appendix I. Hence, the stoichiometric subspace $S = \mathbb{R}^{\mathcal{S}}$. Thus, the rank of the network is equal to the number of species $(s = m)$ as shown in Table 1.

**Table 2.** NRP and STR Network Properties

| Model | | $s = m$ | WR | TM | ET | ILC | PT | SRD | TBD | RES |
|---|---|---|---|---|---|---|---|---|---|---|
| NRP | $\mathcal{N}$ | 40 | NO | NO | NO | NO | YES | YES | YES | YES |
| | $\mathcal{N}_1^*$ | 8 | NO | NO | NO | NO | YES | YES | YES | YES |
| | $\mathcal{N}_2^*$ | 17 | NO | NO | NO | NO | YES | YES | YES | YES |
| | $\mathcal{N}_3^*$ | 15 | NO | NO | NO | NO | YES | YES | YES | YES |
| STR | $\mathcal{N}$ | 37 | NO | NO | X | NO | YES | YES | YES | YES |
| | $\mathcal{N}_1^*$ | 7 | NO | NO | NO | NO | YES | YES | YES | YES |
| | $\mathcal{N}_2^*$ | 14 | NO | NO | X | NO | YES | YES | YES | YES |
| | $\mathcal{N}_3^*$ | 16 | NO | NO | X | NO | YES | YES | YES | YES |

LEGEND: s = m, network rank = number of species ;WR = weakly reversible; TM = t-minimal; ET = endotactic; CC = concordant; ILC = independent linkage classes; PT = point terminal; SRD = sufficient reactant diversity; TBD = terminality bounded by deficiency; RES = the reactant subspace R coincides with the stoichiometric subspace S .

In a weakly reversible (WR) network, $sl = l$. The network numbers of the NRP and STR models in Table 1 show that $sl \neq l$ hence, the networks are not weakly reversible. Since $t \neq l,$ the models are not $t$-minimal (TM).

We used CoNtRol (Donnel et al. 2014), an online toolbox to test the endotacticity (ET) of the network. Due to the complexity of some networks, CoNtRol was not able to generate the said test. That is why the ET column in Table 2 is incomplete. We also used another software tool, ERNEST, consisting of extensive Matlab code which is specifically valuable for large networks such as the NRP and STR to determine the independent linkage class (ILC) property and the network numbers in Table 1. From the report, it showed that the rank of the networks are not equal to the sum of the rank of their corresponding linkage classes. Thus, the “NOs” in the ILC column.

The WR, TM and ET columns in Table 2 are well-studied network properties in the current CRNT literature. Meanwhile, the point terminal (PT) and terminality bounded by deficiency (TBD) columns were introduced in Arceo et al. (2017) as additional network properties in their 15 BST case studies. They defined a network to be point terminal if $t = n - n_r$. For the TBD, we first introduce some convenient notation:

**Definition 16.** The **terminality** $\tau(D)$ (or simply $\tau$) of a digraph D is the non-negative integer $t - l.$

In other words, a $t$-minimal digraph is one with zero terminality and a non- $t$-minimal one with positive terminality.

**Definition 17.** A CRN is of type **terminality bounded by deficiency (TBD)** if $t - l \leq \delta$, otherwise, of type **terminality not deficiency-bounded (TND)**, i.e. $t - l > \delta.$ If it is of type TBD, it is called **terminality properly bounded by deficiency (TPD)** if terminality is less than deficiency and **positive terminality bounded by deficiency (PBD)** if terminality is positive.

A remark of Feliu & Wiuf (2012) that $q < s$ (i.e. low reactant rank) implied degeneracy of all positive equilibria of a mass action system led us to the concept of "sufficient reactant diversity" as a necessary condition for the existence of a non-degenerate equilibrium ($q \geq s \Rightarrow n_r \geq s$).

**Definition 18.** A CRN has **low reactant diversity (LRD)** if $n_r < s,$ otherwise it has **sufficient reactant diversity (SRD)**. An SRD network has **high reactant diversity (HRD)** or **medium reactant diversity (MRD)** if $n_r > s$ or $= s$, respectively.

The next Proposition clarifies the relationship between deficiency-bounded terminality and sufficient reactant diversity.

**Proposition 2.** Let $\mathrm{N}$ be a CRN.

i) A network with deficiency-bounded terminality has sufficient reactant diversity.
ii) If the network is point terminal, then the converse also holds, i.e. TBD ⇐ SRD ( or equivalently TND ⇒ LRD).

**Proof.** From $\delta - \tau(\mathrm{N}) = n - l - s - (t - l) = n - t - s$, we obtain TBD, i.e. $\delta - \tau(\mathrm{N}) \geq 0$ iff $n - t \geq s$. Since $n_r \geq n - t$ for any network, TBD implies SRD. If $\mathrm{N}$ is point terminal $n_r = n - t$, showing that the converse holds. ■

**Corollary 1.** A $t$-minimal network is an SRD network.

**Remark 1.** We have the following ascending chain of networks: reversible ⇒ weakly reversible ⇒ $t$-minimal ⇒ TBD ⇒ SRD.

Since all the S-system embedded networks in Table 2 are PT, Proposition 2.i tells us that the information in the TBD and SRD columns are the same.

In Arceo et al. (2018), a classification of CRNs based on the intersection $R \cap S$ of the reactant and stoichiometric subspaces was introduced. Open networks belong to the network class stoichiometry-determined reactant subspace (SRS) with $R \cap S = R$, since $R \subset S = \mathbb{R}^{\mathcal{S}}$. Table 2 shows that all S-system embedded networks belong to the subset of coincident $R$ and $S$ subspaces, which we denote by RES ($R = S$).

We observe that, for open networks, $R = S$ is equivalent to $q = s$, i.e. the network's rank difference $\Delta(\mathrm{N}) = 0$.

We include the statements of Proposition 2 to collect the relationships between TBD, SRD and RES in the next Theorem and illustrate the relationships for S-system embedded networks in Figure 2.

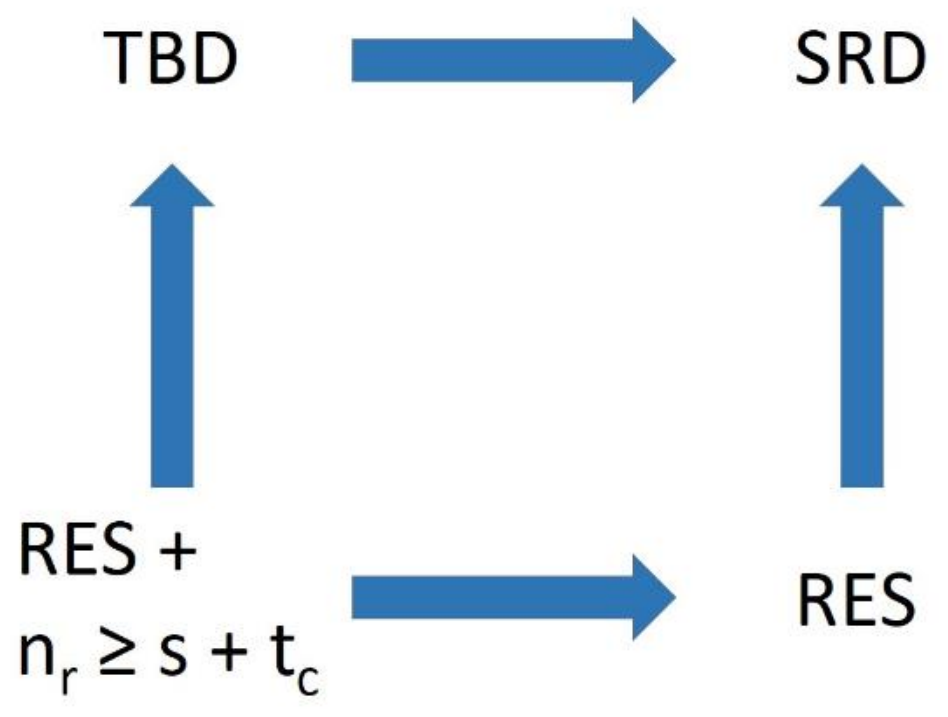

**Figure 2.** Relationships for S-system embedded networks

**Theorem 1.**

i) A TBD network has sufficient reactant diversity. If the network is point terminal, then the converse holds, i.e. TBD ⇔ SRD.

ii) A TPD network has high reactant diversity (HRD). If the network is point terminal, then the converse also holds, i.e. TPD ⇔ HRD.

iii) An RES network has sufficient reactant diversity.

iv) $RES \cap TBD = RES \iff n_r \geq s + t_c$. If the network is point terminal, then RES implies TBD. If the network is not point terminal, then it has high reactant diversity $(n_r > s)$ and reactant deficiency $\delta_p > 0$.

**Proof.** i) This was already shown in Proposition 2.

ii) Given a TPD network, $t - l < \delta \Rightarrow s < n - t \Rightarrow s < n_r - t_c \Rightarrow n_r > s$, since $t_c \geq 0$. If the network is point terminal, $t_c = 0$, and the converse holds.

iii) For any CRN, we have $n_r \geq q$. Since the CRN is RES, $q = s$, which establishes the sufficient reactant diversity.

iv) From $\delta - \tau(\mathrm{N}) = n - t - s = n - (t_c + t_p) - s = n - (t_c + n - n_r) - s = n_r - q - t_c = n_r - (s + t_c)$, we obtain TBD ⇔ $n_r \geq s + t_c$, so that the first claim follows. For point terminal networks $t_c = 0$, and ii) establishes the validity of the right-hand side. If the network is not point terminal, $t_c > 0$, hence $n_r > s$ and $\delta_p := n_r - q = n_r - s > 0$. ■

**Remark 2.** The inequality $n_r \geq s + t_c$ expresses clearly which additional characteristic an RES network needs – beyond SRD – to imply deficiency-bounded terminality.

These network properties of the NRP and STR show a very high degree coincidence with the eight embedded S-systems from BST Case Studies in Arceo et al. (2017) as seen in Table 3.

**Table 3.** Network properties of S-system embedded networks of Arceo et al. (2017).

| Model | $s = m$ | WR | TM | ET | ILC | PT | SRD | TBD | RES |
|---|---|---|---|---|---|---|---|---|---|
| ARL1-S | 17 | NO | NO | NO | NO | YES | YES | YES | YES |
| ARL3-S | 4 | NO | NO | NO | NO | YES | YES | YES | YES |
| ARL4-S | 7 | NO | NO | NO | NO | YES | YES | YES | YES |
| CPA3-S | 9 | NO | NO | NO | NO | YES | YES | YES | YES |
| CPA4-S | 5 | NO | NO | NO | NO | YES | YES | YES | YES |
| ECJ0-S | 16 | NO | NO | NO | NO | YES | YES | YES | YES |
| ECJ2-S | 5 | NO | NO | NO | NO | YES | YES | YES | YES |
| ERM0-S | 5 | NO | NO | NO | NO | YES | YES | YES | YES |

LEGEND: s = m, network rank = number of species ;WR = weakly reversible; TM = t-minimal; ET = endotactic; CC = concordant; ILC = independent linkage classes; PT = point terminal; SRD = sufficient reactant diversity; TBD = terminality bounded by deficiency; RES = the reactant subspace R coincides with the stoichiometric subspace S .

## 3.2 Concordance, weakly monotonic kinetics and non-inhibitory Power Law Kinetics

We consider a reaction network $(\mathcal{S}, \mathcal{C}, \mathcal{R})$ with stoichiometric subspace $S \subset \mathbb{R}^{\mathcal{S}}$ and we let $L: \mathbb{R}^{\mathcal{R}} \to S$ be the linear map defined by

$$L\alpha = \sum_{y \to y' \in \mathcal{R}} \alpha_{y \to y'}(y' - y) \qquad (1)$$

Note that the real scalar multipliers $\{\alpha_{y \to y'}\}_{y \to y' \in \mathcal{R}}$ in equation (1) are permitted to be positive, negative or zero.

**Definition 19.** The reaction network $(\mathcal{S}, \mathcal{C}, \mathcal{R})$ is **concordant** if there do not exist an $\alpha \in ker\ L$ and a nonzero $\sigma \in S$ having the following properties:

i) For each $y \to y' \in \mathcal{R}$ such that $\alpha_{y \to y'} \neq 0$, $supp\ y$ contains a species $s$ for which $sgn\ \sigma_s = sgn\ \alpha_{y \to y'}$.

ii) For each $y \to y' \in \mathcal{R}$ such that $\alpha_{y \to y'} = 0$, either $\sigma_s = 0$ for all $s \in supp\ y$ or else $supp\ y$ contains a species $s$ and $s'$ for which $sgn\ \sigma_s = -sgn\ \sigma_{s'}$, both not zero.

A network that is not concordant is **discordant.**

The set of weakly monotonic kinetics is defined as follows:

**Definition 20.** A kinetics $K$ for a reaction network $(\mathcal{S},\mathcal{C},\mathcal{R})$ is **weakly monotonic** if, for each pair of compositions $c^*$ and $c^{**}$, the following implications hold for each reaction $y \to y' \in \mathcal{R}$ such that $supp\ y \subset supp\ c^*$ and $supp\ y \subset supp\ c^{**}$.

i) $K_{y\to y'}(c^{**}) > K_{y\to y'}(c^*) \Rightarrow$ there is a species $s \in supp\ y$ with $c_s^{**} > c_s^*$.

ii) $K_{y\to y'}(c^{**}) = K_{y\to y'}(c^*) \Rightarrow c_s^{**} > c_s^*$ for all $s \in supp\ y$ or else there are species $s, s' \in supp\ y$ with $c_s^{**} > c_s^*$ and $c_{s'}^{**} > c_{s'}^*$.

The final concept we need, to formulate the basic result, is that of non-inhibitory Power Law kinetics:

**Definition 21.** A power law kinetics is in PL-NIK (N) (or has **non-inhibitory kinetics**) if

i) the kinetic order matrix $F$ is non-negative and

ii) a kinetic order $f_{r,s} > 0$ iff the species $s$ is an element of $supp\ \rho(r)$.

**Theorem 2.** (Shinar & Feinberg 2012). For a CRN $(\mathcal{S},\mathcal{C},\mathcal{R})$, the following conditions are equivalent:

i) The network is concordant.

ii) The network is injective against all weakly monotonic kinetics.

iii) The network is injective against all non-inhibitory power law kinetics.

We now present one of the main results of this paper to prove discordance of S-system networks.

**Theorem 3.**

i) An S-system reaction network with two or more dependent species is discordant.

ii) A non-constant S-system reaction network with one dependent species is discordant if $A \neq 0$ and the product of its rate constants ($\alpha$ and $\beta$) and kinetic orders ($g$ and $h$) $\alpha\beta gh \neq 0$, otherwise concordant.

**Proof.** i) To show discordance, we only need to construct a PL-NIK with a non-injective SFRF on the reaction network. We consider the total representation of the S-system first. The SFRF of any PL kinetics maybe decomposed into an $f_D$ (containing all coordinate functions for dependent species) and $f_i$ with the coordinate functions for the independent species. Since each $f_{i,j}$ has the form $f_{i,j} = a_j > 0$, it is clear that the SFRF is injective iff $f_D$ is injective. We will construct a PL-NIK whose $f_D$ is multistationary, implying that it is not injective. To achieve this, we choose the kinetic order matrix as follows:

- The kinetics is PL-NIK.
- The first column of the derived $A$ matrix consists only of zeros, i.e. $g_{i1} = h_{i1}$, for all $i = 1, \dots, m$.
- $A \neq 0$ (this is possible since the system has two or more dependent species).

We have an S-system with $\det A = 0$, which according to the above, either has no steady state or infinitely many steady states. Since the rate constants can be arbitrarily chosen from the positive real numbers, we adjust them so that we have the latter alternative. Since the network is open, this shows that $f_D$ is not injective in the single stoichiometric class.

ii) We begin this time with $A = 0$. Since we assumed it is not constant, then $\alpha \neq \beta$. So, the kinetics $= (\alpha - \beta)X^g = (\alpha - \beta \exp(g \ln X))$ is strictly monotonic, and hence injective for any positive g. Hence, the CRNs are concordant according to Theorem 2. If $A$ is nonzero, which is equivalent to $\det A$ nonzero, the system has a unique positive equilibrium. If one of the rate constants or the kinetic orders is $0$, then we obtain a function as in the $A = 0$ case plus a constant. Hence, the system is also injective. To show that it is not injective for both rate constants and both kinetic orders nonzero, we need to look at its derivative. Since the S-system in one variable has the form $f(x) = \alpha X^g - \beta X^h$ and $f'(x) = \alpha g X^{g-1} - \beta h X^{h-1}$. One easily checks that $f'(x) < (=)(>) 0$ iff $x^{g-h} < (=)(>) \frac{\beta h}{\alpha g}$. From Calculus, it follows that the function is strictly decreasing and strictly increasing left and right of the particular $x$-value. The value is between its steady state at $0$ and its positive steady state at $\frac{\beta}{\alpha}$, since $g > h$ (i.e. production > degradation). This means that $f$ is not injective in the interval $(0, \frac{\alpha}{\beta})$, so that it is discordant according to Theorem 2. A similar argument holds for $h > g$, the function is not injective in the interval $(0, \frac{\beta}{\alpha})$. For the constant case, the same argument as in (i) applies.

These considerations also cover the embedded representation case, since the same $f_D$ describes the embedded system, by considering the powers of the independent species as part of the rate constants. ■

Using this theorem, the NRP and STR $\mathcal{N}$ and their embedded networks $\mathcal{N}_i^*$ are discordant since all the species are dependent.

## 4. THE SUBNETWORKS OF $\mathcal{N}$ AND MODULARITY

The reaction set partition of the full network $\mathcal{N}$ also induces a decomposition into $k$ subnetworks $\mathcal{N}_1, \ldots, \mathcal{N}_k$. These subnetworks are not CRNs of S-systems since they have species which are neither dependent nor independent variables. However, they are quite similar, and, in particular $\mathcal{N}_i$ has rank $m_i$ of the embedded networks $\mathcal{N}_i^*$. There are also more species in $\mathcal{N}_i$ than in $\mathcal{N}_i^*$ because of an addition of “foreign species” (those not in the subsystem).

**Table 4.** NRP and STR network numbers for $\mathcal{N}$ and its subnetworks $\mathcal{N}_1, \mathcal{N}_2$ and $\mathcal{N}_3$.

| Network numbers | NRP | | | | STR | | | |
|---|---|---|---|---|---|---|---|---|
| | $\mathcal{N}$ | $\mathcal{N}_1$ | $\mathcal{N}_2$ | $\mathcal{N}_3$ | $\mathcal{N}$ | $\mathcal{N}_1$ | $\mathcal{N}_2$ | $\mathcal{N}_3$ |
| Species ($\boldsymbol{m}$) | 40 | 11 | 18 | 17 | 37 | 9 | 19 | 19 |
| Irreversible species ($\boldsymbol{m_{rev}}$) | 0 | 0 | 0 | 0 | 1 | 0 | 1 | 0 |
| Complex ($\boldsymbol{n}$) | 98 | 21 | 36 | 43 | 95 | 19 | 33 | 46 |
| Source complex ($\boldsymbol{n_r}$) | 60 | 12 | 20 | 28 | 60 | 12 | 19 | 29 |
| Linkage class ($\boldsymbol{l}$) | 19 | 5 | 3 | 13 | 22 | 5 | 6 | 14 |
| Strong linkage class ($\boldsymbol{sl}$) | 98 | 21 | 36 | 43 | 94 | 19 | 32 | 46 |
| Terminal linkage class ($\boldsymbol{t}$) | 38 | 9 | 16 | 15 | 35 | 7 | 13 | 17 |
| Rank ($\boldsymbol{s}$) | 40 | 8 | 17 | 15 | 37 | 7 | 14 | 16 |
| Reactant rank ($\boldsymbol{q}$) | 40 | 11 | 18 | 17 | 37 | 9 | 19 | 19 |
| Deficiency ($\boldsymbol{\delta}$) | 39 | 8 | 16 | 15 | 36 | 7 | 13 | 15 |
| Reactant deficiency ($\boldsymbol{\delta_p}$) | 20 | 1 | 2 | 11 | 23 | 3 | 0 | 10 |

In terms of the network properties, both the subnetworks and the embedded networks have the same network properties (these properties are similar to those presented in Table 2). We now have the next basic fact:

**Proposition 3.** The subnetworks $\mathcal{N}_i$ constitute an independent decomposition of $\mathcal{N}$.

**Proof:** The $\mathcal{N}_i$ are subnetworks of $\mathcal{N}$ induced by the partition of the reaction set. Since their ranks sum up to the rank of $\mathcal{N}$, the decomposition is independent. ■

We now relate the subnetworks $\mathcal{N}_i$ to the embedded networks $\mathcal{N}_i^*$ of $\mathcal{N}$. For each $i$, there is a digraph homomorphism from $\mathcal{N}_i$ to $\mathcal{N}_i^*$. A digraph homomorphism maps the vertices while preserving adjacency. For a complex in $\mathcal{N}_i$, its image is its “projection” to $\mathcal{N}_i^*$, i.e. the terms with common species are left out.

**Proposition 4.** The map $\phi: \mathcal{N}_i \rightarrow \mathcal{N}_i^*$ is a digraph homomorphism. Each such map is surjective, and if injective, is an isomorphism of the subdigraphs.

**Proof:** The construction of $i^{th}$ embedded network involves removing species not in the species set $\mathcal{S}_i$ from the complexes of $\mathcal{N}_i$. Hence, there is a surjective map $\phi: \mathcal{R}_i \rightarrow \mathcal{R}_i^*$. Since any complex is a reactant or product of a reaction, one obtains a map $\phi: \mathcal{C}_i \rightarrow \mathcal{C}_i^*$ too. This map is clearly a digraph homomorphism. ■

The subnetworks $\mathcal{N}_i$ is surjective digraph homomorphic to the embedded networks $\mathcal{N}_i^*$ in both the NRP and STR models. Moreover, the subnetworks $\mathcal{N}_2$ ($\mathcal{N}_1$) is injective to the embedded network $\mathcal{N}_2^*(\mathcal{N}_1^*)$ in NRP (STR).

### 4.1 Modularity of digraph divisions and modularity of CRN decompositions

Biological systems often display an organization into functional modules and hence, it is important that models of such systems capture characteristics of the modular structure. Modularity was initially introduced by Newman and Girvan for the case of undirected networks while Arenas proposed an extension of this for directed networks (Li & Schuurmans 2011). Their extension is based on the observation that the existence of a directed edge $(i, j)$ between nodes $i$ and $j$, depends on the out-degree and in-degree of nodes $i$ and $j$ respectively. The modularity for directed network, denoted by $Q$, is expressed as

$$Q = \frac{1}{A} \sum_{i,j} \left[ A_{i,j} - \frac{k_i^{out}\, k_j^{in}}{A} \right] \delta\,(c_i, c_j)$$

where $A$ is the total number of arcs in the network, $A_{i,j}$ is the number of arcs from $i$ to $j$, $k_i^{out}$ and $k_j^{in}$ are the outdegree and indegree of the nodes $i$ and $j$, respectively and $\delta(c_i, c_j) = 1$ (i.e., if nodes $i$ and $j$ belong to the same module) and $0$, otherwise.

In terms of the structure or graph of a network, modularity is designed to measure the strength of division of a network into modules (clusters or communities). Good divisions, which have high modularity values, are those with dense edge connections between the vertices within a module but sparse connections between vertices in different modules (Li & Schuurmans 2011).

A systems biologist describes modules from a graph-theoretical point of view as a group of nodes that are more strongly intraconnected than interconnected while a geneticist might consider a set of co-expressed or co-regulated genes a module (Lorenz et al. 2011)

When we compute the modularities $Q$ of the divisions of the digraphs in Figure 1 and the reaction graph of the CRN decompositions of $\mathcal{N}$ (which is also a digraph) for NRP and STR, we obtain the following surprising results (shown in Table 5):

**Table 5.** Modularity $Q$ of the divisions of the digraph and the reaction graph of $\mathcal{N}$ for NRP and STR.

| Model | $Q$ (digraph) | $Q$ (reaction graph) |
|---|---|---|
| NRP | 0.4405 | 0.3217 |
| STR | 0.4472 | 0.1989 |

Paradoxically, the digraph model indicates that STR is more modular than NRP, while the same measure applied to the reaction graph of the CRN indicates the opposite. Our conclusion from this computation is that because a CRN has a richer structure than just a digraph, one needs to modify or expand the concept of modularity for a CRN to include aspects of its stoichiometric structure. In the following, we introduce an initial concept which is admittedly specific for S-system CRNs but might lead in the right direction.

Now if two subnetworks (i.e. subdigraphs) are in different connected components, then they are "physically" isolated, and hence, the question of modularity is trivial. We hence assume in the following that the subdigraphs in question are in the same connected component.

Any arc in the digraph (biochemical map) whose source and target vertices lie in different subdigraphs lead to the occurrence of common species between the two subnetworks of $\mathcal{N}$ in the S-system CRN's. Since by assumption, the digraph under consideration is connected, for each subdigraph, there is at least one such arc and hence at least one common species with another subnetwork. Based on this, we introduce the concept of species coupling level of a subnetwork:

**Definition 22.** The **species coupling level** $c_S(N')$ of a subnetwork $N'$ in $N$ is the ratio of the number of occurrences of common species in the reactant and product complexes of the input and output reactions, respectively and the number of occurrences of non-common species in the reactant complexes of the input reactions.

With respect to the digraph of the S-system, the numerator counts the number of arcs between vertices from different subdigraphs coming into the subdigraph while the denominator counts the number of arcs between vertices within the subdigraph.

**Remark 3.** The species coupling level of a network is equal to the sum of the species coupling levels of its subnetworks.

Table 6 compares the modularity of the digraphs with the species coupling levels for NRP and STR.

**Table 6.** Modularity $Q$ of the digraphs and the species coupling levels $\sum c_S(\mathcal{N}_i)$ for NRP and STR.

| Model | $Q$ (digraph) | $\sum c_S(\mathcal{N}_i)$ |
|---|---|---|
| NRP | 0.4405 | 0.6236 |
| STR | 0.4472 | 0.5289 |

Since modularity is inversely related to species coupling, we see that the values of the latter for NRP and STR are qualitatively consistent with the modularity values for the corresponding digraphs. In our view, this indicates that stoichiometric level information is useful for modularity considerations.

The NRP and STR decompositions contain only mono-species common complexes. We counted the single non-zero one in the species-level coupling calculation and considered the common zero complex as "outside of the system" studied. We would view the occurrence of common multi-species complexes in other examples as further indicators on the stoichiometric level of lower modularity. We hope to explore these and related concepts in a more general context of developing an appropriate modularity concept for CRN decompositions.

## 5. CONCLUSION

Motivated by Magombedze and Mulder (2013) approach in representing and analyzing the gene regulatory based system model of *Mtb* in modular form. The system partitions the vertices into subsets that form corresponding subsystems.  In this study, these subsystems are treated as embedded networks. In the embedded CRN representation of S-system, it is defined as a species subset and a reaction subset induced by the set of digraph vertices of the subsystem.

In this study, we have

- demonstrated the relationships among the predominant properties of embedded networks of S-system (Proposition 2 and Theorem 1)
- presented the discordance of S-system CRN with at least two dependent species. However, for a non-constant S-system CRN with one dependent species, it is said to be discordant if $A \neq 0$ and the product of its rate constants and kinetic orders $\alpha\beta gh \neq 0$; otherwise it is said to be concordant (Theorem 3).

- shown that the species subsets-induced decomposition of the network induces surjective digraph homomorphisms between the corresponding subnetworks and embedded netwroks (Proposition 4)
- illustrated that the modularity of the decomposition of the S-system is inconsistent with the modularity of the original digraph model of the gene regulatory system. In order for the modularity concept from digraph to capture the stoichiometric structure of CRN, we have introduced the concept of species coupling level for the CRN decompositions.

Biological systems are complex networks that exhibit the orchestrated interplay of a large array of components (Kim 2003). It is common practice for modelers to decompose the complex system into subsystems. It is believed that studying the dynamics and functionality of the subsystems would facilitate understanding of the system as a whole. Henceforth, discovering and analyzing such subsystems are crucial in gaining better understanding of the complex systems (Kim 2003). There are still a lot to explore and study regarding the structural properties with respect to decomposition/separability of subsystems.

## ACKNOWLEDGMENTS

Farinas acknowledges the support of the Commision on Higher Education (CHED) for the CHED-SEGS Scholarship Grant. Lao held research fellowship from De La Salle University.

APPENDIX I:

CRNs of the full S-system and the embedded networks for NRP and STR

**NRP $\mathcal{N}$ full network**

R1: $X_2 \to X_2 + X_1$
R2: $X_1 \to 0$
R3: $X_1 \to X_1 + X_2$
R4: $X_2 \to 0$
R5: $X_1 + X_{13} + X_7 + X_6 + X_4 + X_{16} + X_{11} + X_{38} \to X_1 + X_{13} + X_7 + X_6 + X_4 + X_{16} + X_{11} + X_{38} + X_3$
R6: $X_3 \to 0$
R7: $X_3 + X_{11} + X_6 + X_{12} + X_9 + X_5 + X_{10} \to X_3 + X_{11} + X_6 + X_{12} + X_9 + X_5 + X_{10} + X_4$
R8: $X_4 \to 0$
R9: $X_4 + X_6 + X_9 + X_{11} + X_{12} \to X_4 + X_6 + X_9 + X_{11} + X_{12} + X_5$
R10: $X_5 \to 0$
R11: $X_4 + X_5 + X_9 + X_3 + X_{11} + X_{12} + X_{10} \to X_4 + X_5 + X_9 + X_3 + X_{11} + X_{12} + X_{10} + X_6$
R12: $X_6 \to 0$
R13: $X_3 + X_8 + X_{40} \to X_3 + X_8 + X_{40} + X_7$
R14: $X_7 \to 0$
R15: $X_1 + X_7 + X_{21} + X_{22} + X_{24} + X_{15} \to X_1 + X_7 + X_{21} + X_{22} + X_{24} + X_{15} + X_8$
R16: $X_8 \to 0$
R17: $X_5 + X_4 + X_6 + X_{10} + X_{11} + X_{12} \to X_5 + X_4 + X_6 + X_{10} + X_{11} + X_{12} + X_9$
R18: $X_9 \to 0$
R19: $X_4 + X_6 + X_9 \to X_4 + X_6 + X_9 + X_{10}$
R20: $X_{10} \to 0$
R21: $X_3 + X_4 + X_5 + X_6 + X_9 + X_{12} + X_{17} + X_{13} \to X_3 + X_4 + X_5 + X_6 + X_9 + X_{12} + X_{17} + X_{13} + X_{11}$
R22: $X_{11} \to 0$
R23: $X_4 + X_5 + X_6 + X_9 + X_{11} + X_{17} + X_{13} \to X_4 + X_5 + X_6 + X_9 + X_{11} + X_{17} + X_{13} + X_{12}$
R24: $X_{12} \to 0$
R25: $X_3 + X_1 + X_{11} + X_{18} + X_{16} + X_{36} \to X_3 + X_1 + X_{11} + X_{18} + X_{16} + X_{36} + X_{13}$
R26: $X_{13} \to 0$
R27: $X_1 + X_2 + X_{15} \to X_1 + X_2 + X_{15} + X_{14}$
R28: $X_{14} \to 0$
R29: $X_1 + X_8 + X_{14} \to X_1 + X_8 + X_{14} + X_{15}$
R30: $X_{15} \to 0$
R31: $X_1 + X_3 + X_{13} \to X_1 + X_3 + X_{13} + X_{16}$
R32: $X_{16} \to 0$
R33: $X_{11} + X_{12} \to X_{11} + X_{12} + X_{17}$
R34: $X_{17} \to 0$
R35: $X_{13} + X_{36} \to X_{13} + X_{36} + X_{18}$
R36: $X_{18} \to 0$
R37: $X_7 \to X_7 + X_{19}$
R38: $X_{19} \to 0$
R39: $X_7 \to X_7 + X_{20}$
R40: $X_{20} \to 0$
R41: $X_8 \to X_8 + X_{21}$
R42: $X_{21} \to 0$
R43: $X_1 + X_8 \to X_1 + X_8 + X_{21}$

R44: $X_{22} \rightarrow 0$
R45: $X_1 \rightarrow X_1 + X_{23}$
R46: $X_{23} \rightarrow 0$
R47: $X_1 + X_8 \rightarrow X_1 + X_8 + X_{24}$
R48: $X_{24} \rightarrow 0$
R49: $X_1 \rightarrow X_1 + X_{25}$
R50: $X_{25} \rightarrow 0$
R51: $X_1 \rightarrow X_1 + X_{26}$
R52: $X_{26} \rightarrow 0$
R53: $X_1 + X_2 \rightarrow X_1 + X_2 + X_{27}$
R54: $X_{27} \rightarrow 0$
R55: $X_1 + X_2 \rightarrow X_1 + X_2 + X_{28}$
R56: $X_{28} \rightarrow 0$
R57: $X_2 \rightarrow X_2 + X_{29}$
R58: $X_{29} \rightarrow 0$
R59: $X_2 \rightarrow X_2 + X_{30}$
R60: $X_{30} \rightarrow 0$
R61: $X_2 \rightarrow X_2 + X_{31}$
R62: $X_{31} \rightarrow 0$
R63: $X_2 \rightarrow X_2 + X_{32}$
R64: $X_{32} \rightarrow 0$
R65: $X_2 \rightarrow X_2 + X_{33}$
R66: $X_{33} \rightarrow 0$
R67: $X_2 \rightarrow X_2 + X_{34}$
R68: $X_{34} \rightarrow 0$
R69: $X_2 \rightarrow X_2 + X_{35}$
R70: $X_{35} \rightarrow 0$
R71: $X_1 + X_{13} + X_{18} \rightarrow X_1 + X_{13} + X_{18} + X_{36}$
R72: $X_{36} \rightarrow 0$
R73: $X_2 \rightarrow X_2 + X_{37}$
R74: $X_{37} \rightarrow 0$
R75: $X_3 \rightarrow X_3 + X_{38}$
R76: $X_{38} \rightarrow 0$
R77: $X_4 \rightarrow X_4 + X_{39}$
R78: $X_{39} \rightarrow 0$
R79: $X_1 + X_7 \rightarrow X_1 + X_7 + X_{40}$
R80: $X_{40} \rightarrow 0$

## NRP embedded network $\boldsymbol{\mathcal{N}_1^*}$

R13*: $X_8 + X_{40} \rightarrow X_8 + X_{40} + X_7$
R14: $X_7 \rightarrow 0$
R15*: $X_7 + X_{21} + X_{22} + X_{24} \rightarrow X_7 + X_{21} + X_{22} + X_{24} + X_8$
R16: $X_8 \rightarrow 0$
R37: $X_7 \rightarrow X_7 + X_{19}$
R38: $X_{19} \rightarrow 0$
R39: $X_7 \rightarrow X_7 + X_{20}$
R40: $X_{20} \rightarrow 0$
R41: $X_8 \rightarrow X_8 + X_{21}$

R42: $X_{21} \to 0$
R43*: $X_8 \to X_8 + X_{21}$
R44: $X_{22} \to 0$
R47*: $X_8 \to X_8 + X_{24}$
R48: $X_{24} \to 0$
R79*: $X_7 \to X_7 + X_{40}$
R80: $X_{40} \to 0$

## NRP embedded network $\mathcal{N}_2^*$

R1: $X_2 \to X_2 + X_1$
R2: $X_1 \to 0$
R3: $X_1 \to X_1 + X_2$
R4: $X_2 \to 0$
R27: $X_1 + X_2 + X_{15} \to X_1 + X_2 + X_{15} + X_{14}$
R28: $X_{14} \to 0$
R29*: $X_1 + X_{14} \to X_1 + X_{14} + X_{15}$
R30: $X_{15} \to 0$
R45: $X_1 \to X_1 + X_{23}$
R46: $X_{23} \to 0$
R49: $X_1 \to X_1 + X_{25}$
R50: $X_{25} \to 0$
R51: $X_1 \to X_1 + X_{26}$
R52: $X_{26} \to 0$
R53: $X_1 + X_2 \to X_1 + X_2 + X_{27}$
R54: $X_{27} \to 0$
R55: $X_1 + X_2 \to X_1 + X_2 + X_{28}$
R56: $X_{28} \to 0$
R57: $X_2 \to X_2 + X_{29}$
R58: $X_{29} \to 0$
R59: $X_2 \to X_2 + X_{30}$
R60: $X_{30} \to 0$
R61: $X_2 \to X_2 + X_{31}$
R62: $X_{31} \to 0$
R63: $X_2 \to X_2 + X_{32}$
R64: $X_{32} \to 0$
R65: $X_2 \to X_2 + X_{33}$
R66: $X_{33} \to 0$
R67: $X_2 \to X_2 + X_{34}$
R68: $X_{34} \to 0$
R69: $X_2 \to X_2 + X_{35}$
R70: $X_{35} \to 0$
R73: $X_2 \to X_2 + X_{37}$
R74: $X_{37} \to 0$
R75: $X_3 \to X_3 + X_{38}$
R76: $X_{38} \to 0$
R77: $X_4 \to X_4 + X_{39}$
R78: $X_{39} \to 0$
R79: $X_1 + X_7 \to X_1 + X_7 + X_{40}$

R80: $X_{40} \to 0$

### NRP embedded network $\mathcal{N}_3^*$

R5*: $X_{13} + X_6 + X_4 + X_{16} + X_{11} + X_{38} \to X_{13} + X_6 + X_4 + X_{16} + X_{11} + X_{38} + X_3$
R6: $X_3 \to 0$
R7: $X_3 + X_{11} + X_6 + X_{12} + X_9 + X_5 + X_{10} \to X_3 + X_{11} + X_6 + X_{12} + X_9 + X_5 + X_{10} + X_4$
R8: $X_4 \to 0$
R9: $X_4 + X_6 + X_9 + X_{11} + X_{12} \to X_4 + X_6 + X_9 + X_{11} + X_{12} + X_5$
R10: $X_5 \to 0$
R11: $X_4 + X_5 + X_9 + X_3 + X_{11} + X_{12} + X_{10} \to X_4 + X_5 + X_9 + X_3 + X_{11} + X_{12} + X_{10} + X_6$
R12: $X_6 \to 0$
R17: $X_5 + X_4 + X_6 + X_{10} + X_{11} + X_{12} \to X_5 + X_4 + X_6 + X_{10} + X_{11} + X_{12} + X_9$
R18: $X_9 \to 0$
R19: $X_4 + X_6 + X_9 \to X_4 + X_6 + X_9 + X_{10}$
R20: $X_{10} \to 0$
R21: $X_3 + X_4 + X_5 + X_6 + X_9 + X_{12} + X_{17} + X_{13} \to X_3 + X_4 + X_5 + X_6 + X_9 + X_{12} + X_{17} + X_{13} + X_{11}$
R22: $X_{11} \to 0$
R23: $X_4 + X_5 + X_6 + X_9 + X_{11} + X_{17} + X_{13} \to X_4 + X_5 + X_6 + X_9 + X_{11} + X_{17} + X_{13} + X_{12}$
R24: $X_{12} \to 0$
R25*: $X_3 + X_{11} + X_{16} + X_{36} \to X_3 + X_{11} + X_{16} + X_{36} + X_{13}$
R26: $X_{13} \to 0$
R31*: $X_3 + X_{13} \to X_3 + X_{13} + X_{16}$
R32: $X_{16} \to 0$
R33: $X_{11} + X_{12} \to X_{11} + X_{12} + X_{17}$
R34: $X_{17} \to 0$
R35: $X_{13} + X_{36} \to X_{13} + X_{36} + X_{18}$
R36: $X_{18} \to 0$
R71*: $X_{13} + X_{18} \to X_{13} + X_{18} + X_{36}$
R72: $X_{36} \to 0$
R75: $X_3 \to X_3 + X_{38}$
R76: $X_{38} \to 0$
R77: $X_4 \to X_4 + X_{39}$
R78: $X_{39} \to 0$

### STR $\mathcal{N}$ full network

R1: $0 \to Y_1$
R2: $Y_1 \to 0$
R3: $Y_{11} + Y_{32} + Y_6 + Y_5 + Y_4 + Y_7 + Y_{19} \to Y_{11} + Y_{32} + Y_6 + Y_5 + Y_4 + Y_7 + Y_{19} + Y_2$
R4: $Y_2 \to 0$
R5: $Y_{10} \to Y_{10} + Y_3$
R6: $Y_3 \to 0$
R7: $Y_6 + Y_5 + Y_7 + Y_2 + Y_8 + Y_9 \to Y_6 + Y_5 + Y_7 + Y_2 + Y_8 + Y_9 + Y_4$
R8: $Y_4 \to 0$
R9: $Y_2 + Y_6 + Y_4 + Y_7 + Y_{12} + Y_{13} \to Y_2 + Y_6 + Y_4 + Y_7 + Y_{12} + Y_{13} + Y_5$
R10: $Y_5 \to 0$
R11: $Y_2 \to Y_2 + Y_6$

R12: $Y_6 \rightarrow 0$
R13: $Y_6 + Y_3 + Y_2 + Y_5 + Y_4 + Y_8 + Y_9 \rightarrow Y_6 + Y_3 + Y_2 + Y_5 + Y_4 + Y_8 + Y_9 + Y_7$
R14: $Y_7 \rightarrow 0$
R15: $Y_3 + Y_{15} + Y_{14} + Y_9 + Y_4 + Y_7 \rightarrow Y_3 + Y_{15} + Y_{14} + Y_9 + Y_4 + Y_7 + Y_8$
R16: $Y_8 \rightarrow 0$
R17: $Y_3 + Y_{14} + Y_8 + Y_7 + Y_4 \rightarrow Y_3 + Y_{14} + Y_8 + Y_7 + Y_4 + Y_9$
R18: $Y_9 \rightarrow 0$
R19: $Y_{17} + Y_3 + Y_1 + Y_{32} \rightarrow Y_{17} + Y_3 + Y_1 + Y_{32} + Y_{10}$
R20: $Y_{10} \rightarrow 0$
R21: $Y_{32} + Y_{21} + Y_6 + Y_2 + Y_{19} + Y_{20} \rightarrow Y_{32} + Y_{21} + Y_6 + Y_2 + Y_{19} + Y_{20} + Y_{11}$
R22: $Y_{11} \rightarrow 0$
R23: $Y_5 + Y_{29} + Y_{18} + Y_{30} + Y_{31} + Y_{13} + Y_{33} \rightarrow Y_5 + Y_{29} + Y_{18} + Y_{30} + Y_{31} + Y_{13} + Y_{33} + Y_{12}$
R24: $Y_{12} \rightarrow 0$
R25: $Y_{12} + Y_5 + Y_{23} + Y_{31} + Y_{18} + Y_{30} \rightarrow Y_{12} + Y_5 + Y_{23} + Y_{31} + Y_{18} + Y_{30} + Y_{13}$
R26: $Y_{13} \rightarrow 0$
R27: $Y_3 + Y_8 + Y_9 \rightarrow Y_3 + Y_8 + Y_9 + Y_{14}$
R28: $Y_{14} \rightarrow 0$
R29: $Y_3 + Y_8 \rightarrow Y_3 + Y_8 + Y_{15}$
R30: $Y_{15} \rightarrow 0$
R31: $Y_1 + Y_{10} \rightarrow Y_1 + Y_{10} + Y_{16}$
R32: $Y_{16} \rightarrow 0$
R33: $Y_1 + Y_{10} \rightarrow Y_1 + Y_{10} + Y_{17}$
R34: $Y_{17} \rightarrow 0$
R35: $Y_{12} + Y_{13} + Y_{30} \rightarrow Y_{12} + Y_{13} + Y_{30} + Y_{18}$
R36: $Y_{18} \rightarrow 0$
R37: $Y_2 + Y_{11} \rightarrow Y_2 + Y_{11} + Y_{19}$
R38: $Y_{19} \rightarrow 0$
R39: $Y_1 + Y_{11} \rightarrow Y_1 + Y_{11} + Y_{20}$
R40: $Y_{20} \rightarrow 0$
R41: $Y_{11} + Y_{32} + Y_2 \rightarrow Y_{11} + Y_{32} + Y_2 + Y_{21}$
R42: $Y_{21} \rightarrow 0$
R43: $Y_1 + Y_{10} \rightarrow Y_1 + Y_{10} + Y_{22}$
R44: $Y_{22} \rightarrow 0$
R45: $Y_1 + Y_{13} \rightarrow Y_1 + Y_{13} + Y_{23}$
R46: $Y_{23} \rightarrow 0$
R47: $Y_1 \rightarrow Y_1 + Y_{24}$
R48: $Y_{24} \rightarrow 0$
R49: $Y_1 \rightarrow Y_1 + Y_{25}$
R50: $Y_{25} \rightarrow 0$
R51: $Y_3 + Y_8 + Y_9 \rightarrow Y_3 + Y_8 + Y_9 + Y_{26}$
R52: $Y_{26} \rightarrow 0$
R53: $Y_6 \rightarrow Y_6 + Y_{27}$
R54: $Y_{27} \rightarrow 0$
R55: $Y_{10} + Y_6 \rightarrow Y_{10} + Y_6 + Y_{28}$
R56: $Y_{28} \rightarrow 0$
R57: $Y_{12} \rightarrow Y_{12} + Y_{29}$
R58: $Y_{29} \rightarrow 0$
R59: $Y_{12} + Y_{13} + Y_{18} \rightarrow Y_{12} + Y_{13} + Y_{18} + Y_{30}$
R60: $Y_{30} \rightarrow 0$
R61: $Y_{12} + Y_{13} \rightarrow Y_{12} + Y_{13} + Y_{31}$

R62: $Y_{31} \rightarrow 0$
R63: $Y_2 + Y_3 + Y_{10} + Y_{11} + Y_{21} \rightarrow Y_2 + Y_3 + Y_{10} + Y_{11} + Y_{21} + Y_{32}$
R64: $Y_{32} \rightarrow 0$
R65: $Y_{12} \rightarrow Y_{12} + Y_{33}$
R66: $Y_{33} \rightarrow 0$
R67: $Y_1 \rightarrow Y_1 + Y_{34}$
R68: $Y_{34} \rightarrow 0$
R69: $Y_1 \rightarrow Y_1 + Y_{35}$
R70: $Y_{35} \rightarrow 0$
R71: $Y_1 \rightarrow Y_1 + Y_{36}$
R72: $Y_{36} \rightarrow 0$
R73: $Y_1 \rightarrow Y_1 + Y_{37}$
R74: $Y_{37} \rightarrow 0$

**STR embedded network $\mathcal{N}_1^*$**

R23*: $Y_{29} + Y_{18} + Y_{30} + Y_{31} + Y_{13} + Y_{33} \rightarrow Y_{29} + Y_{18} + Y_{30} + Y_{31} + Y_{13} + Y_{33} + Y_{12}$
R24: $Y_{12} \rightarrow 0$
R25: $Y_{12} + Y_{31} + Y_{18} + Y_{30} \rightarrow Y_{12} + Y_{31} + Y_{18} + Y_{30} + Y_{13}$
R26: $Y_{13} \rightarrow 0$
R35: $Y_{12} + Y_{13} + Y_{30} \rightarrow Y_{12} + Y_{13} + Y_{30} + Y_{18}$
R36: $Y_{18} \rightarrow 0$
R57: $Y_{12} \rightarrow Y_{12} + Y_{29}$
R58: $Y_{29} \rightarrow 0$
R59: $Y_{12} + Y_{13} + Y_{18} \rightarrow Y_{12} + Y_{13} + Y_{18} + Y_{30}$
R60: $Y_{30} \rightarrow 0$
R61: $Y_{12} + Y_{13} \rightarrow Y_{12} + Y_{13} + Y_{31}$
R62: $Y_{31} \rightarrow 0$
R65: $Y_{12} \rightarrow Y_{12} + Y_{33}$
R66: $Y_{33} \rightarrow 0$

**STR embedded network $\mathcal{N}_2^*$**

R1: $0 \rightarrow Y_1$
R2: $Y_1 \rightarrow 0$
R19*: $Y_{17} + Y_1 \rightarrow Y_{17} + Y_1 + Y_{10}$
R20: $Y_{10} \rightarrow 0$
R31: $Y_1 + Y_{10} \rightarrow Y_1 + Y_{10} + Y_{16}$
R32: $Y_{16} \rightarrow 0$
R33: $Y_1 + Y_{10} \rightarrow Y_1 + Y_{10} + Y_{17}$
R34: $Y_{17} \rightarrow 0$
R39*: $Y_1 \rightarrow Y_1 + Y_{20}$
R40: $Y_{20} \rightarrow 0$
R43: $Y_1 + Y_{10} \rightarrow Y_1 + Y_{10} + Y_{22}$
R44: $Y_{22} \rightarrow 0$
R45*: $Y_1 \rightarrow Y_1 + Y_{23}$
R46: $Y_{23} \rightarrow 0$
R47: $Y_1 \rightarrow Y_1 + Y_{24}$

R48: $Y_{24} \rightarrow 0$
R49: $Y_1 \rightarrow Y_1 + Y_{25}$
R50: $Y_{25} \rightarrow 0$
R55*: $Y_{10} \rightarrow Y_{10} + Y_{28}$
R56: $Y_{28} \rightarrow 0$
R67: $Y_1 \rightarrow Y_1 + Y_{34}$
R68: $Y_{34} \rightarrow 0$
R69: $Y_1 \rightarrow Y_1 + Y_{35}$
R70: $Y_{35} \rightarrow 0$
R71: $Y_1 \rightarrow Y_1 + Y_{36}$
R72: $Y_{36} \rightarrow 0$

**STR embedded network $\mathcal{N}_3^*$**

R3: $Y_{11} + Y_{32} + Y_6 + Y_5 + Y_4 + Y_7 + Y_{19} \rightarrow Y_{11} + Y_{32} + Y_6 + Y_5 + Y_4 + Y_7 + Y_{19} + Y_2$
R4: $Y_2 \rightarrow 0$
R5*: $0 \rightarrow Y_3$
R6: $Y_3 \rightarrow 0$
R7: $Y_6 + Y_5 + Y_7 + Y_2 + Y_8 + Y_9 \rightarrow Y_6 + Y_5 + Y_7 + Y_2 + Y_8 + Y_9 + Y_4$
R8: $Y_4 \rightarrow 0$
R9*: $Y_2 + Y_6 + Y_4 + Y_7 \rightarrow Y_2 + Y_6 + Y_4 + Y_7 + Y_5$
R10: $Y_5 \rightarrow 0$
R11: $Y_2 \rightarrow Y_2 + Y_6$
R12: $Y_6 \rightarrow 0$
R13: $Y_6 + Y_3 + Y_2 + Y_5 + Y_4 + Y_8 + Y_9 \rightarrow Y_6 + Y_3 + Y_2 + Y_5 + Y_4 + Y_8 + Y_9 + Y_7$
R14: $Y_7 \rightarrow 0$
R15: $Y_3 + Y_{15} + Y_{14} + Y_9 + Y_4 + Y_7 \rightarrow Y_3 + Y_{15} + Y_{14} + Y_9 + Y_4 + Y_7 + Y_8$
R16: $Y_8 \rightarrow 0$
R17: $Y_3 + Y_{14} + Y_8 + Y_7 + Y_4 \rightarrow Y_3 + Y_{14} + Y_8 + Y_7 + Y_4 + Y_9$
R18: $Y_9 \rightarrow 0$
R21: $Y_{32} + Y_{21} + Y_6 + Y_2 + Y_{19} + Y_{20} \rightarrow Y_{32} + Y_{21} + Y_6 + Y_2 + Y_{19} + Y_{20} + Y_{11}$
R22: $Y_{11} \rightarrow 0$
R27: $Y_3 + Y_8 + Y_9 \rightarrow Y_3 + Y_8 + Y_9 + Y_{14}$
R28: $Y_{14} \rightarrow 0$
R29: $Y_3 + Y_8 \rightarrow Y_3 + Y_8 + Y_{15}$
R30: $Y_{15} \rightarrow 0$
R37: $Y_2 + Y_{11} \rightarrow Y_2 + Y_{11} + Y_{19}$
R38: $Y_{19} \rightarrow 0$
R41: $Y_{11} + Y_{32} + Y_2 \rightarrow Y_{11} + Y_{32} + Y_2 + Y_{21}$
R42: $Y_{21} \rightarrow 0$
R51: $Y_3 + Y_8 + Y_9 \rightarrow Y_3 + Y_8 + Y_9 + Y_{26}$
R52: $Y_{26} \rightarrow 0$
R53: $Y_6 \rightarrow Y_6 + Y_{27}$
R54: $Y_{27} \rightarrow 0$
R63*: $Y_2 + Y_3 + Y_{11} + Y_{21} \rightarrow Y_2 + Y_3 + Y_{11} + Y_{21} + Y_{32}$
R64: $Y_{32} \rightarrow 0$